\documentclass{article}
\usepackage{slashed,epsfig,amsmath,amssymb,enumitem}
\usepackage[papersize={8.5in,11in}]{geometry}
\usepackage{graphicx}
\newcommand{\be}{\begin{equation}}
\newcommand{\ee}{\end{equation}}
\begin{document}
\title{Model of Boson and Fermion Particle Masses}
\author{J. W. Moffat\\~\\
Perimeter Institute for Theoretical Physics, Waterloo, Ontario N2L 2Y5, Canada\\
and\\
Department of Physics and Astronomy, University of Waterloo, Waterloo,\\
Ontario N2L 3G1, Canada}
\maketitle




\begin{abstract}
The boson and fermion particle masses are calculated in a finite quantum field theory. The field theory satisfies Poincar\'e invariance, unitarity and microscopic causality, and all loop graphs are finite to all orders of perturbation theory. The infinite derivative nonlocal field interactions are regularized with a mass (length) scale parameter $\Lambda_i$. The $W$, $Z$ and Higgs boson masses are calculated from finite one-loop self-energy graphs. The $W^{\pm}$ mass is predicted to be $M_W=80.05$ GeV, and the higher order radiative corrections to the Higgs boson mass $m_H=125$ GeV are damped out above the regulating mass scale parameter $\Lambda_H=1.57$ TeV. The three generations of quark and lepton masses are calculated from finite one-loop self-interactions, and there is an exponential spacing in mass between the quarks and leptons.
\end{abstract}
\maketitle

\section{Introduction}

A paramount mystery in particle physics is the origin of the three families of quark and lepton masses. The standard model (SM) of quarks and leptons interacting with gauge forces leads to a successful quantitative theory. The Higgs field is coupled to the quarks and leptons through the Yukawa interactions, and their different interaction strengths with the Higgs doublet describe their masses. However, the SM does not provide a fundamental understanding of the bizarre pattern of quark and lepton masses that exhibit 12 orders of magnitude differences in masses. There are 22 free parameters associated with the particle flavors and 28 free parameters in the Majorana extended SM, questioning whether the SM is a fundamental theory of nature.

Most attempts to solve the flavor puzzle of SM involve enlarging the group structure $SU(3)_C\times S(2)_L\times U(1)_Y$ that describes the strong, weak and electromagnetic forces. It has been proposed that the three families are distinguished by new quantum numbers belonging to the flavors, $G_{fl}$, extending the SM group structure to $G_{fl}\times SU(3)_C\times S(2)_L\times U(1)_Y$. Recently, Weinberg~\cite{Weinberg} has pursued the idea that only the third generation of quarks and leptons masses are generated through the spontaneous symmetry breaking of the vacuum and by tree graphs, while the second and third generation masses are generated by radiative loop graphs. Weinberg extended the standard model gauge group by $SO(3)_L\times SO(3)_R$ with the three  generations of quarks and charged leptons furnishing the separate representations $(1,3)$. Nine doublets of scalar Higgs bosons with accompanying gauge vector bosons and additional new free parameters are required to produce a renormalizable model. The new undetected particles are elevated to a higher energy. He concluded that this scheme does not lead to realistic models of the masses.

We will pursue an alternative solution of the particle masses mystery by employing the finite quantum field theory (finite QFT)~\cite{Moffat1989,Moffat1990,MoffatWoodard1991,Moffat1991,ClaytonMoffat1991,MoffatRobbins1991,MoffatHand1991,WoodardKleppe1991,WoodardKleppe1992,Hand1992,Cornish1992a,Cornish1992b,
Cornish1992c,WoodardKleppe1993,ClaytonDemopolous1994,Paris1995,Troost1996,Moffat1998,Moffat2007,MoffatToth2010,Moffat2020}. The QFT realizes a Poincar\'e invariant, unitary QFT and all quantum loop graphs are ultraviolet finite to all orders of perturbations theory. Although the field operators and the interactions of particles are nonlocal, the model satisfies microscopic causality~\cite{Moffat2020}. The SM $SU(2)\times U(1)$ electroweak sector has been formulated as spontaneous breaking of the gauge invariant sector with masses produced by the symmetry breaking~\cite{Higgs1964,Brout1964,Kibble1964,Higgs1966,Kibble1967,Halzen,Burgess}. The finite QFT model allows for an alternative interpretation of the elecroweak $SU(2)\times U(1)$ sector. Because the loop graphs are finite an infinite renormalization of particle interactions is not required. This allows for finite field theory interactions with massive bosons and fermions in the Lagrangian. The idea that the $SU(2)\times U(1)$ Lagrangian has initially to be massless at the outset to guarantee a gauge invariant and renormalizable scheme is discarded. Moreover, the assumption that the classical Higgs potential has the form:
\be
\label{SMpotential}
{\cal L}_\phi=-\mu^2\phi^\dagger\phi+\partial_\mu\phi^\dagger\partial^\mu\phi+\lambda(\phi^\dagger\phi)^2,
\ee
where $\phi$ is the complex Higgs field need not be made. The boson and fermion masses are calculated from perturbative one-loop graphs with an associated QFT length (mass) scale $\Lambda_i$. By determining the scale $\Lambda_{WZ}$ for the Z boson mass, the mass of the W boson is predicted to be $M_W=80.05$ GeV close to the observed mass $M_W=80.379\pm 0.012$ GeV. Because the Higgs field self-interaction coupling parameter $\lambda$ in the electroweak symmetry broken SM and the Higgs field vacuum expectation value $v=\langle 0|\phi|0\rangle=246$ GeV are discarded, there is one less free parameter in the present particle model. All the low energy predicted decay products and particle productions verified by the LHC will be retained in the alternative model. However, the only electroweak true vacuum will be $v=0$ predicting a stable vacuum in contrast to the standard prediction by the Higgs mechanism of an unstable vacuum at very high energies~\cite{Kusenko1997,Bednyakov2015,Moffat2020}.

A key experimental verification of the standard electroweak symmetry breaking mechanism is to measure the triple and quartic self-interaction terms in the potential $V(\phi)$ after symmetry breaking:
\be
\label{Higgspotential}
V(\phi)=\frac{1}{2}m_H^2\phi^2+\lambda v\phi^3+\lambda\phi^4.
\ee
In the SM the assumed specific self-coupling term after symmetry breaking, $\lambda v\phi^3$, predicts the relation between the self-coupling constant $\lambda$ and the vacuum expectation value $v$, $M_H=\sqrt{2\lambda}v$. The vacuum expectation value $v=246$ GeV is known from the Fermi theory to a precision of 7 digits, leading to the predicted value $\lambda=0.13$ verified experimentally to several digits. A modification of the Higgs potential (\ref{Higgspotential}) will modify these predictions. Only a direct measurement of the Higgs field self-coupling interaction can verify the standard electroweak symmetry breaking model. Until now no measurement of the Higgs self-interaction has been obtained. Indeed, it is not known whether the Higgs field does undergo a self-interaction! A measurement of the second term in the Higgs potential, $\lambda v\phi^3$, can be achieved by measuring the trilinear self-coupling of the Higgs field. This can be done by the detection of a pair of Higgs particles in the final state. This is difficult to accomplish, because the production rate is more than a thousand times smaller than the production of a single Higgs boson~\cite{Carvalho2020}. A detection of the quadrilinear Higgs field self-interaction is well beyond the ability of present day and any near future accelerators. The best hope to detect the Higgs self-interaction lies with $e^+e^-$ colliders such as the Compact Linear Collider (CLIC) which could reach the necessary detection threshold greater than 500 GeV.

\section{Electroweak Model Lagrangian}

We shall concentrate on the electroweak sector of our model. The electroweak model Lagrangian is
\begin{align}
\label{Lagrangian}
{\cal L}_{\rm EW}=\sum_{\psi_L}\tilde\psi_L\biggl[\gamma^\mu\biggl(i\partial_\mu - \frac{1}{2}{\tilde g}\tau^aW^a_\mu - {\tilde g}'\frac{Y}{2}B_\mu\biggr)\biggr]\psi_L
+\sum_{\psi_R}\tilde\psi_R\biggl[\gamma^\mu\biggl(i\partial_\mu - {\tilde g}'\frac{Y}{2}B_\mu\biggr)\biggr]\psi_R\nonumber\\
 -\frac{1}{4}B^{\mu\nu}B_{\mu\nu}
-\frac{1}{4}W_{\mu\nu}^aW^{a\mu\nu} + {\cal L}_M + {\cal L}_{m_f}.
\end{align}
The ${\tau}'s$ are the usual Pauli spin matrices and $\psi_L$ denotes a left-handed fermion (lepton or quark) doublet, and the $\psi_R$ denotes a right-handed fermion singlet. The fermion fields (leptons and quarks) have been written as $SU_L(2)$ doublets and U(1)$_Y$ singlets, and we have suppressed the fermion generation indices. We have $\psi_{L,R}=P_{L,R}\psi$, where $P_{L,R}=\frac{1}{2}(1\mp\gamma_5)$.
The Lagrangian for the neutral scalar Higgs boson is given by
\be
{\cal L}_{\rm Higgs}=\biggl|\biggr(i\partial_\mu -\frac{1}{2}{\tilde g_H}\tau^a W_\mu^a-{\tilde g_H}'\frac{Y}{2}B_\mu\biggr)\phi\biggr|^2
+\frac{1}{2}m_H^2\phi^2,
\ee
where $\phi$ is the isoscalar neutral Higgs field. The photon-fermion Lagrangian is
\be
L_{\rm QED}=\sum_{\psi_L}\tilde\psi_L\biggl[\gamma^\mu\biggl(i\partial_\mu - \frac{1}{2}{\tilde e}\biggr)A_\mu\biggr]\psi_L
+\sum_{\psi_R}\psi_R\biggl[\gamma^\mu\biggl(i\partial_\mu - \frac{1}{2}{\tilde e}\biggr)A_\mu\biggr]\psi_R -\frac{1}{4}F^{\mu\nu}F_{\mu\nu}+{\cal L}_{m_f},
\ee
where
\begin{equation}
\label{Bequation}
B_{\mu\nu}=\partial_\mu B_\nu-\partial_\nu B_\mu,
\end{equation}
\begin{equation}
W^a_{\mu\nu}=\partial_\mu W_\nu^a-\partial_\nu W_\mu^a-{\tilde g}f^{abc}W_\mu^bW_\nu^c,
\end{equation}
and
\be
F_{\mu\nu}=\partial_\mu A_\nu-\partial_\nu A_\mu.
\ee
The Lagrangian for the vector boson mass terms is
\be
{\cal L}_M=\frac{1}{2}M^2_WW^{a\mu} W^a_\mu + \frac{1}{2}M^2_BB^\mu B_\mu,
\ee
and the fermion mass Lagrangian is
\be
\label{fermionmass}
{\cal L}_{m_f}=-\sum_{\psi_L^i,\psi_R^j}m_{ij}^f(\tilde\psi_L^i\psi_R^j + \tilde\psi_R^i\psi_L^j),
\ee
where $M_W$, $M_B$ and $m_{ij}^f$ denote the boson and fermion masses, respectively. Eq.(\ref{fermionmass}) can incorporate massive neutrinos and their flavor oscillations. The mass Lagrangians explicitly break $SU(2)_L\times U(1)_Y$ gauge symmetry.

The $A_\mu$ and $Z_\mu$ are linear combinations of the two fields
$W_{3\mu}$ and $B_\mu$:
\begin{equation}
\label{A} A_\mu=c_wB_\mu+s_wW_{3\mu},
\end{equation}
\begin{equation}
\label{Z} Z_\mu=-s_wB_\mu+c_wW_{3\mu},
\end{equation}
where $c_w=\cos\theta_w$, $s_w=\sin\theta_w$ and the angle $\theta_w$ denotes the weak mixing angle. The
electroweak coupling constants $g$ and $g'$ are related to the
electric charge $e$ by the standard equation
\be
gs_w=g'c_w=e
\ee
and we use the standard normalization $c_w=g/(g^2+g^{'2})^{1/2}$ and $g'/g=\tan\theta_w$.

The fermion and boson fields are local fields and we have
\be
{\tilde g}(p^2)=g{\cal E}(p^2/\Lambda_i^2),
\ee
\be
{\tilde g}'(p^2)=g'{\cal E}(p^2/\Lambda_i^2),
\ee
\be
{\tilde e}(p^2)=e{\cal E}(p^2/\Lambda_i^2).
\ee
\be
{\tilde g_H}(p^2)=g_H{\cal E}(p^2/\Lambda_H^2),
\ee
\be
{\tilde g'_H}(p^2)=g'_H{\cal E}(p^2/\Lambda_H^2).
\ee

We define the entire function distribution operator ${\cal E}$ in terms of the kinetic operator ${\cal K}$:
\be
\label{Edistr}
{\cal E}=\exp\bigg(\frac{{\cal K}}{2\Lambda_i^2}\biggr).
\ee
The Feynman rules for the finite QFT follow as extensions of the local standard QFT. Every internal line in a Feynman diagram can be connected to a regulated propagator:
\be
\label{propagatorReg}
i\tilde\Delta=\frac{i{\cal E}^2}{{\cal K}}=i\int\frac{d\tau}{\Lambda_i^2}\exp\biggl(\tau\frac{\cal K}{\Lambda_i^2}\biggr),
\ee
where we have used the Schwinger proper time method to determine the propagator.

An additional auxiliary propagator was introduced in the formulation of finite QED~\cite{MoffatWoodard1991}:
\be
\label{auxpropagator}
-i\hat\Delta=\frac{i(1-{\cal E}^2)}{{\cal K}}=-i\int_0^1\frac{d\tau}{\Lambda_i^2}\exp\biggl(\tau\frac{{\cal K}}{\Lambda_i^2}\biggr).
\ee
The auxiliary propagator $\hat\Delta$ does not possess poles and does not have particles. Tree order amplitudes such as Compton scattering amplitudes are identical to their local QFT counterparts. The tree amplitudes such as Compton amplitudes are the sum of (\ref{propagatorReg}) and (\ref{auxpropagator}), and this sum gives the standard local propagator and tree graphs and they are free of unphysical couplings.

\section{Microscopic Causality}

The choice of nonlocal interaction coupling distributions can be equivalently described by constant coupling constants $g$, $g'$, $e$ and $g_H$ and nonlocal field operators defined by
\be
\tilde\phi(x)=\int d^4x'{\cal F}(x-x')\phi(x')={\cal F}({\Box_x}(x))\phi(x),
\ee
where $\phi(x)$ is the local field operator and $\Box_x=\partial^\mu\partial_\mu$. The local field operator $\phi(x)$ satisfies the commutation relation:
\be
\bar\Delta(x-x')=i[\phi(x),\phi(x')],
\ee
where
\be
\bar\Delta(x)=i\int\frac{d^4k}{(2\pi)^4}\exp(-ik\cdot (x-x'))\epsilon(k^0)2\pi\delta(k^2-m^2),
\ee
is the Pauli-Jordan propagator and
\be
\epsilon(k^0)=\theta(k^0)-\theta(-k^0)=1\, {\rm for}\, k^0 > 0\, {\rm and}\,-1\, {\rm for}\, k^0 < 0,
\ee
and
\be
\theta(k^0)=1\, {\rm for}\, k^0 > 0\, {\rm and}\, 0\, {\rm for}\, k^0 < 0.
\ee
We have
\be
\bar\Delta(x)=\frac{1}{2\pi}\epsilon(x^0)\delta(x^2)-\frac{m}{4\pi\sqrt{(x^2)}}\theta(x^2)\epsilon(x^0)J_1(m\sqrt{(x^2)}),
\ee
where $J_1$ is a Bessel function. The Pauli-Jordan propagator satisfies outside the light cone:
\be
\bar\Delta(x-x')=0,\quad (x-x')^2 < 0,
\ee
and
\be
[\phi(x),\phi(x')]=0,\quad (x-x')^2 < 0.
\ee

We can now prove that the finite QFT satisfies microscopic causality~\cite{Moffat2020,Ogarkov}. We define
\be
{\cal F}(x-x')={\cal F}(\Box_x)\delta^4(x-x'),
\ee
where
\be
\tilde\phi(x)=\int d^4x'{\cal F}(x-x')\phi(k)\exp(-ik\cdot x)=\int\frac{d^4k}{(2\pi)^4}\phi(k){\cal F}(-k^2)\exp(-ik\cdot x).
\ee
We can now calculate the commutator:
\be
[\tilde\phi(x),\tilde\phi(x')]=[{\cal F}(\Box_x)\phi(x),{\cal F}(\Box_{x'})\phi(x')]={\cal F}(\Box_x){\cal F}(\Box_{x'})[\phi(x),\phi(x')].
\ee
This leads to
\begin{align}
[\tilde\phi(x),\tilde\phi(x')]={\cal F}(\Box_x){\cal F}(\Box_{x'})\int\frac{d^4k}{(2\pi)^4}\exp(-ik\cdot (x-x'))\epsilon(k^0)2\pi\delta(k^2-m^2)\nonumber\\
=\int\frac{d^4k}{(2\pi^4)}{\cal F}^2(-m^2)\exp(-ik\cdot (x-x'))\epsilon(k^0)2\pi\delta(k^2-m^2)
={\cal F}^2(-m^2)(-i)\bar\Delta(x-x'),
\end{align}
It then follows that for the nonlocal field operator $\tilde\phi(x)$:
\be
[\tilde\phi(x),\tilde\phi(x')]=0,\quad (x-x')^2 < 0.
\ee
This proves that the finite QFT satisfies microscopic causality.

\section{The Boson Masses}

The vector propagator for the scattering of longitudinally polarized vector bosons is
\be
iD^{\mu\nu}(p^2)=\frac{-i\eta^{\mu\nu}}{p^2-\Pi_{Vf}^T(p^2)},
\ee
where $\eta_{\mu\nu}$ denotes the Minkowski spacetime metric, and we explicitly indicated the dependence of the self-energy and the propagator on momentum. This differs from the vector boson propagator of the SM in that the squared mass $m_V^2$ of the vector boson is replaced by the self-energy term $\Pi_{Vf}^T$. For an on-shell vector boson, demanding agreement with the standard model requires that the following consistency equation be satisfied:
\be
m_V^2=\Pi_{Vf}^T(m_V^2).
\label{eq:vecmass}
\ee

In the transverse sector for the $Z-Z$ part, we get~\cite{Moffat2007,MoffatToth2010}:
\be
-i\Pi_{Zf}^T=-\frac{1}{2}\frac{i(g^2+g'^2)\Lambda_{WZ}^2}{(4\pi)^2}\sum_\psi[(K_{m_1m_2}-L_{m_1m_2})+P_{m_1m_2}(2c_w^4+s_w^432(Q-T_3)^2-16s_w^2c_w^2T^3(Q-T_33))].
\label{eq:ZZPi}
\ee
where we have summed over fermion loop doublets and masses $m_1$ and $m_2$. The pure photon sector gives
\be
-i\Pi_{Af}^T=-\frac{1}{2}\frac{i(g^2+g'^2)\Lambda_{WZ}^2}{(4\pi)^2}c_w^2s_w^2\sum_\psi P_{m_1m_2}[2+32(Q-T_3)^2+16T_3(Q-T_33)].
\label{eq:photon}
\ee
Here, $Q=T_3+\frac{1}{2}Y$ is the electric charge and $T_3$ is the third component of isospin and we have
\be
\label{eq:P}
P_{m_1m_2}=-\frac{p_E^2}{\Lambda_W^2}\int_0^{\frac{1}{2}}d\tau\tau(1-\tau)\left[E_1\left(\tau\frac{p_E^2}{\Lambda_W^2}+f_{m_1m_2}\right)
+E_1\left(\tau\frac{p_E^2}{\Lambda_W^2}+f_{m_2m_1}\right)\right],
\ee
\be
K_{m_1m_2}=\int_0^{\frac{1}{2}}d\tau(1-\tau)\left[\exp\left(-\tau\frac{p_E^2}{\Lambda_W^2}-f_{m_1m_2}\right)
+\exp\left(-\tau\frac{p_E^2}{\Lambda_W^2}-f_{m_2m_1}\right)\right],
\ee
\be
L_{m_1m_2}=\int_0^{\frac{1}{2}}d\tau(1-\tau)\left[f_{m_1m_2}E_1\left(\tau\frac{p_E^2}{\Lambda_{WZ}^2}+f_{m_1m_2}\right)
+f_{m_2m_1}E_1\left(\tau\frac{p_E^2}{\Lambda_{WZ}^2}+f_{m_2m_1}\right)\right],
\ee
where
\be
f_{m_1m_2}=\frac{m_1^2}{\Lambda_{WZ}^2}+\frac{\tau}{1-\tau}\frac{m_2^2}{\Lambda_{WZ}^2}.
\ee
Moreover, $E_1$ is the exponential integral:
\be
E_1(z)\equiv\int^\infty_z dy\frac{\exp(-y)}{y}=-\ln(z)-\gamma_e-\sum^\infty_{n=1}\frac{(-z)^n}{nn!},
\ee
and $\gamma_e$ is Euler's constant. If we insert this into the quadratic terms in the action and invert, we get the corrected propagator (in a general gauge):
\be
iD^{\mu\nu}=-i\left(\frac{\eta^{\mu\nu}-\frac{p^\mu p^\nu}{p^2}}{p^2-\Pi_{Vf}^T}+\frac{\frac{\xi p^\mu p^\nu}{p^2}}{p^2-\xi\Pi_{Vf}^L}\right),
\label{eq:vecprop}
\ee
and when the longitudinal piece is nonzero in the unitary gauge (where only the physical particle spectrum remains), we have no unphysical poles in the longitudinal sector. In this way, we can assure ourselves that we are not introducing spurious degrees of freedom into the theory.

In the diagonalized $W^\pm$ sector, we get
\begin{align}
-i\Pi_{W^\pm f}^L=&-\frac{ig^2\Lambda_{WZ}^2}{(4\pi)^2}\sum_{q^L}(K_{m_1m_2}-L_{m_1m_2}),\\
-i\Pi_{W^\pm f}^T=&-\frac{ig^2\Lambda_{WZ}^2}{(4\pi)^2}\sum_{q^L}(K_{m_1m_2}-L_{m_1m_2}+2P_{m_1m_2}).
\label{eq:WWPi}
\end{align}
The pure photon sector gives
\be
-i\Pi_{Af}^T=-\frac{1}{2}\frac{i(g^2+g'^2)\Lambda_{WZ}^2}{(4\pi)^2}c_w^2s_w^2\sum_\psi P_{m_1m_2}[2+32(Q-T^3)^2+16T^3(Q-T^3)].
\label{eq:photon}
\ee
We observe from (\ref{eq:P}) and (\ref{eq:photon}) that $\Pi_{Af}^T(0)=0$, guaranteeing a massless photon.

We obtain for the mixing sector:
\be
-i\Pi_{AZf}^T=-\frac{1}{2}\frac{i(g^2+g'^2)\Lambda_{WZ}^2}{(4\pi)^2}c_w^2s_w^2
\sum_\psi P_{m_1m_2}[2c_w^2-32s_w^2(Q-T_3)^2-16T_3(Q-T_3)(s_w^2-c_w^2)].
\ee

To calculate boson masses, we note the form of the massive vector boson propagator (\ref{eq:vecprop}). When we consider the scattering of longitudinally polarized vector bosons, the terms containing $p_\mu p_\nu$ cancel out. In the remaining term, $\Pi_{Vf}^T$ appears in the same place where, in the SM, $M_V^2$ is present. We therefore make the identification
\begin{equation}
M_V^2=\Pi_{Vf}^T.
\end{equation}
This allows us to calculate the masses of the $W^\pm$ and $Z^0$ bosons or conversely, use their experimentally known masses to calculate $\Lambda_{WZ}$.

For the $Z$-boson, the on-shell mass $M_Z$ is determined by the right-hand side of (\ref{eq:vecmass}) and by (\ref{eq:ZZPi}), and we find that it contains terms that include the electroweak coupling constant, the Weinberg angle, fermion masses, and the $\Lambda_{WZ}$ parameter. As all these except $\Lambda_{WZ}$ are known from experiment, the equation
\be
M_Z^2=\Pi^T_{Zf}(M_Z^2),
\ee
the right-hand side of which contains $\Lambda_{WZ}$ through (\ref{eq:ZZPi}), can be used to determine $\Lambda_{WZ}$. Using the values
$g=0.649$, $s_w=0.2312$, and $m_t=172.76$ GeV, and noting that the calculation is not sensitive to the much smaller masses of the other 11 fermions, we get
\be
\Lambda_{WZ}=542~\mathrm{GeV},
\ee
where the precision of $\Lambda_{WZ}$ is determined by the precision to which the $Z$-mass is known, and it is not sensitive to the lack of precision knowledge of the top quark mass or the other quark masses. Knowing $\Lambda_{WZ}$ allows us to solve the consistency equation for the $W$-boson mass. Treating $M_W$ as unknown, we solve using (\ref{eq:WWPi}):
\be
M_W^2=\Pi^T_{Wf}(m_W^2),
\ee
for $M_W$, and obtain
\be
M_W=80.05~\mathrm{GeV}.
\ee
This result, which does not incorporate radiative corrections, is actually slightly closer to the experimental value $M_W=80.379\pm 0.012$~GeV than the comparable tree-level standard model prediction $M_W\simeq 79.95$~GeV. This is anticipated as our regularization scheme will introduce some suppression of higher-order corrections at the energy scale of $M_W$ and $\Lambda_{WZ}$.

We can obtain a non-trivial prediction of the $\rho$ parameter. Using the definition
\be
\rho=\frac{M_W^2}{M_Z^2\cos^2\theta_w},
\ee
we get
\be
\rho\simeq 1.0023.
\ee

The Lagrangian we consider for a real scalar field describing the Higgs boson is given by
\be
{\cal L}_H=\frac{1}{2}\phi\Box\phi-\frac{1}{2}m_H^2\phi^2-\frac{1}{4!}\lambda\phi^4.
\ee

The entire function distribution for the scalar Higgs field is given by
\be
{\cal E}=\exp\biggl(\frac{\Box+m_H^2}{2\Lambda_H^2}\biggr).
\ee
The propagator in Euclidean momentum space using the Schwinger proper time formalism is
\be
i\tilde\Delta_H(p)\equiv \frac{i{\cal E}^2}{p^2+m_H^2}=i\int\frac{d\tau}{\Lambda_H^2}
\exp\biggl[-\tau\biggl(\frac{p^2+m_H^2}{\Lambda_H^2}\biggr)\biggr].
\ee

We assume that the Higgs bare mass $m_{0H}=0$ and identify the dominant one-loop top quark radiative self-energy with the Higgs mass, $m_H\sim \delta m_H$. We obtain
\be
m_H^2=\frac{y_t^2}{16\pi^2}\Lambda_H^2+\delta{\cal O}(m_{\rm weak}^2),
\ee
where $y_t$ is the top quark coupling constant. With the choice $y_t\sim 1$ and $m_H=125$ GeV, we find
\be
\Lambda_H\sim 1.57\, {\rm TeV}.
\ee
The higher order Higgs mass radiative corrections will be damped out at higher energies greater than $\Lambda_H$, because of the exponential damping caused by the entire function in the propagator~\cite{Moffat2020}.

\section{Fermion Masses}

In the standard electroweak model, the fermion masses are generated
through Yukawa couplings. We will generate fermion masses from
the finite one-loop fermion self-energy graph by means of a
Nambu-Jona-Lasinio mechanism~\cite{Nambu}. A fermion particle satisfies
\be
\label{meq} i\slashed{p}+m_{0f}+\Sigma(p)=0,
\ee
for $i\slashed{p}+m_f=0$ where $m_{0f}$ is the bare fermion
mass, $m_f$ is the observed fermion mass and $\Sigma(p)$ is the
finite proper self-energy part. We have
\begin{equation}
\label{sigmaeq}
m_f-m_{0f}=\Sigma(p,m_f,g,\Lambda_f)\vert_{i\slashed{p}+m_f=0}.
\end{equation}
Here, $\Lambda_f$ denote the energy scales for lepton
and quark masses.

The finite self-energy contribution obtained by joining together
two fermion-boson vertices is given by~\cite{ClaytonMoffat1991,MoffatWoodard1991,Moffat2007,MoffatToth2010}:
\be
-i\Sigma_1(p)=\int
\frac{d^4k}{(2\pi)^2}(ig^2_f\gamma^\mu)\biggl(\frac{-i}{\slashed{p}
+m_f-i\epsilon}\biggr)(ig^2_f\gamma^\nu)
\biggl(\frac{-i\eta_{\mu\nu}}{k^2+m^2-i\epsilon}\biggr)
\exp\biggl[-\biggl(\frac{p^2+m_f^2}{\Lambda_f^2}\biggr)
-\biggl(\frac{q^2+m_f^2}{\Lambda_f^2}\biggr)-\frac{k^2}{\Lambda_f^2}\biggr],
\ee
where $g^2_f$ is a fermion coupling constant containing quark
color factors for strong coupling and is a weak coupling constant for leptons and $q=p-k$.  We have neglected the small vector boson mass contribution, $M_V^2/\Lambda_f^2$.

The propagators are now converted to Schwinger integrals and the
momentum integration is performed to give
\be
-i\Sigma_1(p)=-g^2_f\exp\biggl[-\biggl(\frac{p^2+m_f^2}{\Lambda_f^2}\biggr)\biggr]
\int^\infty_1\frac{d\tau_1}{\Lambda_f^2}\int^\infty_1\frac{d\tau_2}{\Lambda_f^2}\int\frac{d^4k}{(2\pi)^4}
(2\slashed{q}+4m_f)\exp\biggl[-\tau_1\biggl(\frac{q^2+m^2}{\Lambda_f^2}\biggr)-\tau_2\frac{k^2}{\Lambda_f^2}\biggr].
\ee
This result can be expressed as
\be
-i\Sigma_1(p)=\frac{-ig_f^2}{8\pi^2}\exp\biggl[-\biggl(\frac{p^2+m_f^2}{\Lambda_f^2}\biggr)\biggr]
\int^\infty_1 d\tau_1\int^\infty_1d\tau_2\biggl[\frac{\tau_2}{(\tau_1+\tau_2)^3}\slashed
{p}+\frac{2m_f}{(\tau_1+\tau_2)^2}\biggr]
\exp\biggl(-\frac{\tau_1\tau_2}{\tau_1+\tau_2}\frac{p^2}{\Lambda_f^2}
-\tau_1\frac{m_f^2}{\Lambda_f^2}\biggr).
\ee
Here, we have performed a rotation to Euclidean momentum space, accounting for the factor of $i$.

Another contribution $\Sigma_2(p)$ to the self-energy comes from
the tadpole fermion-boson self-energy graph:
\be
-i\Sigma_2(p)=\int\frac{d^4k}{(2\pi)^2}(-ig_f^2)\gamma^\mu(\slashed
{q}-m_f)\gamma^\nu\biggl(\frac{-i\eta_{\mu\nu}}{k^2-i\epsilon}\biggr)
\int_0^1\frac{d\tau}{\Lambda_f^2}\exp\biggl[-\biggl(\frac{p^2+m_f^2}{\Lambda_f^2}\biggr)
-\tau\biggl(\frac{q^2+m_f^2}{\Lambda_f^2}\biggr)-\frac{k^2}{\Lambda_f^2}\biggr].
\ee

Adding together the two diagram contributions $\Sigma_1(p)$ and
$\Sigma_2(p)$, we obtain
\be
\Sigma(p)=\frac{g^2_f}{8\pi^2}\exp\biggl[-\biggl(\frac{p^2+m_f^2}{\Lambda_f^2}\biggr)\biggr]
\int^1_0 dx(x\slashed{p}+2m_f)E_1\biggl[(1-x)\frac{p^2}{\Lambda^2_f}+\biggl(\frac{1-x}{x}\biggr)\frac{m_f^2}{\Lambda_f^2}\biggr].
\ee
By developing an asymptotic expansion in $\Lambda_f$ and expanding the exponential integral, we get
\be
\Sigma(p)=\frac{\alpha_f}{2\pi}\biggl[\biggl(\frac{1}{2}\slashed{p}
+2m_f\biggr)\ln(\Lambda_f^2)-\biggl(\frac{1}{2}\slashed{p}+2m_f\biggr)\gamma_e
+\frac{1}{2}\slashed{p}-\int^1_0 dx(x\slashed{p}
+2m_f)\ln(xp^2+m_f^2)+O\biggl[\frac{\ln(\Lambda^2_f)}{\Lambda_f^2}\biggr],
\ee
where $\alpha_f=g^2_f/4\pi$.


The fermion mass is now identified with $\Sigma(p)$ at $p=0$ and we choose $m_{0f}=0$:
\be
m_f=\Sigma(0)=\frac{\alpha_fm_f}{\pi}\biggl[\ln\biggl(\frac{\Lambda_f^2}{m_f^2}\biggr)
-\gamma_e\biggr]
+O\biggl[\frac{\ln(\Lambda^2_f)}{\Lambda_f^2}\biggr].
\ee
This equation has two solutions: either $m_f=0$, or
\be
1=\frac{\alpha_f}{\pi}\biggl[\ln\biggl(\frac{\Lambda_f^2}{m_f^2}\biggr)-\gamma_e\biggr].
\ee
The first trivial solution corresponds to the standard
perturbation result. The second non-trivial solution will
determine $m_f$ in terms of $\alpha_f$ and $\Lambda_f$ and leads
to the fermion ``mass gap'' equation
\be
m_f=\Lambda_f\exp\biggl[-\frac{1}{2}\biggl(\frac{\pi}{\alpha_f}+\gamma_e\biggr)\biggr].
\ee

We observe that the fermion masses reveal an exponential displacement as is suggested by the experimental hierarchical values of the quark and lepton masses.
In Table 1, we display the quark and lepton masses and the corresponding mass scales $\Lambda_f$. We have chosen for the strong coupling constant
$\alpha_s(M_Z)=0.117\pm 0.0007$, where $\alpha_s=g_s^2/4\pi$ is evaluated at the $m_Z$ pole and $g_s=1.213$, and for the electroweak coupling constant $g_w=0.649$ and $\alpha_w=0.0516$.
\begin{table}[h!]
\begin{center}
\caption{Fermion masses and $\Lambda_f$}
\label{tab:table1}
\begin{tabular}{ccc}
\hline
${\rm fermion}$ & ${\rm mass}_f$ & $\Lambda_f$ \\
\hline
t & 172.76 GeV & $1.69\times 10^5$ TeV\\
b & 4.18 GeV & $4.11\times 10^3$ TeV\\
c & 1.28 GeV & $1.16\times 10^3$ TeV\\
s & 0.095 GeV & 86 TeV\\
d & $3.98\times 10^{-3}$ GeV & 3.98 TeV\\
u & $1.87\times 10^{-3}$ GeV & 1.72 TeV\\
$\tau$ & 1.777 GeV & $3.82\times 10^{10}$ TeV\\
$\mu$ & 0.106 GeV & $2.28\times 10^9$ TeV \\
$e$ & $5.11\times 10^{-4}$ GeV & $1.08\times 10^7$ TeV \\
$\nu_{\tau}$ & 0.015 GeV & $3.2\times 10^8$ TeV \\
$\nu_{\mu}$ & $1.9\times 10^{-4}$ GeV & $4.1\times 10^6$ TeV \\
$\nu_e$ & $0.20\times 10^{-8}$ GeV & $43.0$ TeV \\
\hline
\end{tabular}
\end{center}
\end{table}

\section{Conclusions}

We have formulated a particle model based on the finite QFT which satisfies unitarity and microcausality. The quantum loops ar finite to all orders of perturbation theory with mass (length) scales $\Lambda_i$ for the bosons and fermions. The regularizing length scales $\Lambda_i$ determine the lengths at which a particle probe cannot be performed at a localized value of energy. The local point-like determination of particle interactions becomes smeared out and the interactions are no longer described by a Dirac delta function distribution. Instead, they are described by a infinite derivative, entire function distribution ${\cal E}(p^2)$. The perturbative infinite renormalizations of charge and mass in local QFT are replaced by finite renormalizations. It is no longer required that the Lagrangian of the theory satisfies a gauge symmetry as for QED and QCD with their massless photon and gluon, respectively. We adopt the assumption that the electroweak sector $SU(2)\times U(1)$ is a broken symmetry with non-zero $W$ and $Z$ bosons and non-zero quark, lepton and neutrino masses.

The masses of bosons and fermions are determined by their perturbative self-energies. An approximate calculation of the boson one-loop W boson self-energy predicts the W boson mass, $M_W=80.05$ GeV and the Z mass is determined to be its experimental $M_Z=91$ GeV with a mass scale $\Lambda_{WZ}=542$ GeV. The Higgs mass is $m_H=125$ GeV for a mass scale $\Lambda_H=1.57$ TeV and radiative Higgs mass corrections are exponentially damped out for energies greater than $\Lambda_H$. The fermion masses determined by the one-loop perturbative fermion self-energies display a mass spectrum in which the quarks, leptons and neutrinos are exponentially separated in mass, a prediction in agreement with the observed experimental pattern of fermion masses.

\section*{Acknowledgments}

I thank Martin Green and Viktor Toth for helpful discussions. Research at the Perimeter Institute for Theoretical Physics is supported by the Government of Canada through industry Canada and by the Province of Ontario through the Ministry of Research and Innovation (MRI).

\end{document}